\documentclass[twocolumn,amsmath,amssymb,prb]{revtex4-1}
\usepackage{graphicx}
\usepackage{amssymb}

\begin{document}

\bibliographystyle{aip}

\title{Quantifying defects in graphene via Raman spectroscopy at different excitation energies}

\author{L. G. Can\c{c}ado$^{1}$, A. Jorio$^{1}$, E. H. Martins Ferreira$^{2}$, F. Stavale$^{2}$, C. A. Achete$^{2}$, R. B. Capaz$^{3}$,  M. V. O. Moutinho$^{3}$, A. Lombardo$^{4}$, T. Kulmala$^{4}$, and A. C. Ferrari$^{4}$}
\address{$^{1}$Departamento de F\'{\i}sica, Universidade Federal de Minas Gerais, 30123-970, Belo Horizonte, Brazil}
\address{$^{2}$Divis\~{a}o de Metrologia de Materiais, Instituto Nacional de Metrologia, Normaliza\c{c}\~{a}o e Qualidade Industrial (INMETRO),
Duque de Caxias, RJ 25250-020, Brazil,}
\address{$^{3}$Instituto de F\'{\i}sica, Universidade Federal do Rio de Janeiro, Cx. Postal 68528, Rio de Janeiro, 21941-972 RJ, Brazil,}
\address{$^{4}$Department of Engineering, University of Cambridge, Cambridge CB3 0FA, UK.}

\begin{abstract}
We present a Raman study of Ar$^{+}$-bombarded graphene samples with increasing ion doses. This allows us to have a controlled, increasing, amount of defects. We find that the ratio between the D and G peak intensities for a given defect density strongly depends on the laser excitation energy. We quantify this effect and present a simple equation for the determination of the point defect density in graphene via Raman spectroscopy for any visible excitation energy. We note that, for all excitations, the D to G intensity ratio reaches a maximum for an inter-defect distance\,$\sim$\,3\,nm. Thus, a given ratio could correspond to two different defect densities, above or below the maximum. The analysis of the G peak width and its dispersion with excitation energy solves this ambiguity.
\end{abstract}

\maketitle

\section{Introduction}

Quantifying defects in graphene related systems, which include a large family of $sp^{2}$ carbon structures, is crucial both to gain insight in their fundamental properties, and for applications. In graphene, this is a key step towards the understanding of the limits to its ultimate mobility\cite{ni,kim,fuhrer}. Large efforts have been devoted to quantify defects and disorder using Raman spectroscopy for nanographites\cite{tk,ferrari00,cancado06,cancado07,nema,knight,matt,vidano1,sinha,lespade,cuesta,wilhem,ferrari01,ferrarissc,rsbook,pimenta07}, amorphous carbons\cite{ferrari00,ferrari01,ferdian,ach,cn,csi,rsbook}, carbon nanotubes\cite{hulman05,chou07}, and graphene\cite{teweldebrhan09,lucchese10,jorio10,martinsferreira10,ferrari06,ferrarissc,cedge,canedge,krauss2010,beams}. The first attempt was the pioneering work of Tuinstra and Koenig (TK)\cite{tk}. They reported the Raman spectrum of graphite and nano-crystalline graphite, and assigned the mode at $\sim$1580\,cm$^{-1}$ to the high frequency $E_{\rm 2g}$ Raman allowed optical phonon, now known as G peak\cite{vidano1}. In defected and nanocrystalline samples they measured a second peak at $\sim$1350\,cm$^{-1}$, now known as D peak\cite{vidano1}. They assigned it to an $A_{\rm 1g}$ breathing mode at the Brillouin Zone (BZ) boundary \textbf{K}, activated by the relaxation of the Raman fundamental selection rule $\textbf{q}\,\approx\,\textbf{0}$, where $\textbf{q}$ is the phonon wavevector\cite{tk}. They noted that the ratio of the D to G intensities varied inversely with the crystallite size, $L_{\rm a}$. Ref.\cite{ferrari00} noted the failure of the TK relation for high defect densities, and proposed a more complete amorphization trajectory valid to date. Refs.\cite{sinha,matt,ferrari00,ferrari01} reported a significant excitation energy dependence of the intensity ratio. Refs.\cite{cancado06,cancado07} measured this excitation laser energy dependency in the Raman spectra of nanographites, and the ratio between the D and G bands was shown to depend on the fourth power of the excitation laser energy $E_{\rm L}$.

There is, however, a fundamental geometric difference between defects related to the size of a nano-crystallite and point defects in the $sp^{2}$ carbon lattices, resulting in a different intensity ratio dependence on the amount of disorder. Basically, the amount of disorder in a nano-crystallite is given by the amount of border (one-dimensional defects) with respect to the total crystallite area, and this is a measure of the nano-crystallite size $L_{\rm a}$. In graphene with zero-dimensional point-like defects, the distance between defects, $L_{\rm D}$, is a measure of the amount of disorder, and recent experiments show that different approaches must be used to quantify $L_{\rm D}$ and $L_{\rm a}$ by Raman spectroscopy~\cite{lucchese10}. The effect of changing $L_{\rm D}$ on peak width, frequency, intensity, and integrated area for many Raman peaks in single layer graphene was studied in Ref.~\cite{martinsferreira10}, and extended  to N-layer graphene in Ref.~\cite{jorio10}, all using a single laser line  $E_{\rm L}\,=\,2.41\,$eV.

Here, to fully accomplish the protocol for quantifying point-like defects in graphene using Raman spectroscopy (or equivalently, $L_{\rm D}$), we use different excitation laser lines in ion-bombarded samples and measure the D to G peak intensity ratio. This ratio is denoted in literature as $I_{\rm D}/I_{\rm G}$ or I(D)/I(G), while the ratio of their areas, i.e. frequency integrated intensity, as $A_{\rm D}/A_{\rm G}$ or A(D)/A(G). In principle, for small disorder or perturbations, one should always consider the area ratio, since the area under each peak represents the probability of the whole process, considering uncertainty~\cite{martinsferreira10,Basko09a}. However, for large disorder, it is far more informative to decouple the information on peak intensity and full width at half maximum. The latter, denoted in literature as FWHM or $\Gamma$, is a measure of structural disorder\cite{ach,cancado07,martinsferreira10}, while the intensity represents the phonon modes/molecular vibrations involved in the most resonant Raman processes~\cite{ach,ferrari00,ferrari01}. For this reason, in this paper we will consider the decoupled $I_{\rm D}/I_{\rm G}$ and peak widths trends. We find that, for a given $L_{\rm D}$, $I_{\rm D}/I_{\rm G}$ increases as the excitation laser energy increases. We present a set of empirical formulas that can be used to quantify the amount of point-like defects in graphene samples with $L_{\rm D}\,\geq\,$10\,nm  using any excitation laser energy/wavelength in the visible range. The analysis of the D and G peak widths and their dispersions with excitation energy unambiguously discriminate between the two main stages of disordering incurred by such samples.

\section{Results and discussion}

We produce single layer graphene (SLG) samples with increasing defect density by mechanical exfoliation followed by Ar$^{+}$-bombardment, as for the procedure outlined in Ref.\cite{lucchese10}. The ion-bombardment experiments are carried out in an OMICRON VT-STM ultra-high vacuum system (base pressure 5\,$\times$\,10$^{-11}$\,mbar) equipped with an ISE 5 Ion Source. Raman spectra are measured at room temperature with a Renishaw micro-spectrometer. The spot size is $\sim$\,1\,$\mu$m for a 100$\times$ objective, and the power is kept at $\sim$\,1.0\,mW to avoid heating. The excitation energies, $E_{\rm L}$, (wavelengths, $\lambda_{\rm L}$) are: Ti-Sapph 1.58\,eV (785\,nm), He-Ne 1.96\,eV (632.8\,nm), Ar$^+$ 2.41\,eV (514.5\,nm).

\begin{figure}[t!]
\centerline{\includegraphics[width=100mm]{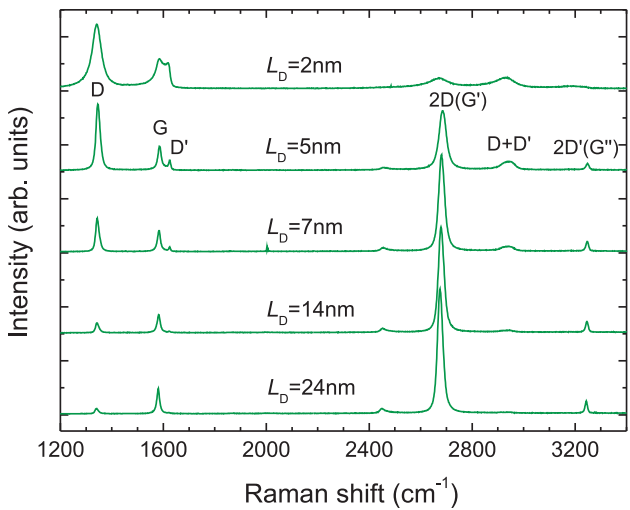}}
\caption[]{Raman spectra of five ion bombarded SLG measured at $E_{\rm L}\,=\,$2.41\,eV ($\lambda_{\rm L}\,=\,$514.5\,nm). The $L_{\rm D}$ values are given according to Ref.~\cite{lucchese10}, and the main peaks are labeled. The notation within parenthesis [e.g. 2D(G$^{\prime}$)] indicate two commonly used notations for the same peak (2D and G$^{\prime}$)~\cite{ferrari06,cancadogprime}.\label{f01}}
\end{figure}

Figure~\ref{f01} plots the Raman spectra of five SLG exposed to different ion bombardment doses in the range 10$^{11}$ Ar$^{+}$/cm$^{2}$ (one defect per 4\,$\times$\,10$^{4}$ C atoms) to 10$^{15}$ Ar$^{+}$/cm$^{2}$ (one defect for every four C atoms). The bombardment procedure described in Ref.~\cite{lucchese10} is accurately reproducible. By tuning the bombardment exposure we generated samples with $L_{\rm D}$\,=\,24,\,14,\,13,\,7,\,5,\,and\,2\,nm. All spectra in Fig. \ref{f01} are taken at $E_{\rm L}\,=\,$2.41\,eV ($\lambda_{\rm L}\,=\,$514.5\,nm).

The Raman spectra in Figure~\ref{f01} consist of a set of distinct peaks. The G and D appear around 1580\,cm$^{-1}$ and 1350\,cm$^{-1}$, respectively. The G peak corresponds to the $E_{\rm 2g}$ phonon at the Brillouin zone center. The D peak is due to the breathing modes of six-atom rings and requires a defect for its activation~\cite{tk,ferrari00,ferrari01,Thomsen00}. It comes from transverse optical (TO) phonons around the \textbf{K} or \textbf{K$^{\prime}$} points in the $1^{\rm st}$ Brillouin zone~\cite{tk,ferrari00,ferrari01}, involves an intervalley double resonance process~\cite{Thomsen00,saitoprl00}, and is strongly dispersive~\cite{vidano} with excitation energy due to a Kohn Anomaly at \textbf{K}\cite{PiscanecPRL}. Double resonance can also happen as intravalley process, i.~e. connecting two points belonging to the same cone around \textbf{K} or \textbf{K$^{\prime}$}~\cite{saitoprl00}. This gives the so-called D$^{\prime}$ peak, which is centered at $\sim~1620$\,cm$^{-1}$ in defected samples measured at 514.5nm\cite{nema}. The 2D peak (also called G$^{\prime}$ in the literature) is the second order of the D peak~\cite{nema,ferrari06}. This is a single peak in single layer graphene, whereas it splits in four in bilayer graphene, reflecting the evolution of the electron band structure~\cite{ferrari06,cancadogprime}. The 2D$^{\prime}$ peak (also called G$^{\prime\prime}$ in analogy to G$^{\prime}$) is the second order of D$^{\prime}$. Since 2D(G$^{\prime}$) and 2D$^{\prime}$(G$^{\prime\prime}$) originate from a process where momentum conservation is satisfied by two phonons with opposite wavevectors, no defects are required for their activation, and are thus always present. On the other hand, the D\,+\,D$^{\prime}$ band ($\sim$\,2940\,cm$^{-1}$) is the combination of phonons with different momenta, around \textbf{K} and $\Gamma$, thus requires a defect for its activation.

Ref.~\cite{ferrari00} proposed a three stage classification of disorder in carbon materials, to simply assess the Raman spectra of carbons along an amorphization trajectory leading from graphite to tetrahedral amorphous carbon: 1) graphite to nanocrystalline graphite; 2) nanocrystalline graphite to low $sp^{3}$ amorphous carbon; 3) low $sp^{3}$ amorphous carbon to high $sp^{3}$ (tetrahedral) amorphous carbon. In the study of graphene, stages 1 and 2 are the most relevant and are summarized here.

In stage 1, the Raman spectrum evolves as follows~\cite{ferrari00,lucchese10,martinsferreira10}: a) D appears and $I_{\rm D}/I_{\rm G}$ increases; b) D$^{\prime}$ appears; c) all peaks broaden. In the case of graphite the D and 2D lose their doublet structure~\cite{cancadolc,ferrari00}; e) D\,+\,D$^{\prime}$ appears; f) at the end of stage 1, G and D$^{\prime}$ are so wide that they start to overlap. If a single lorentzian is used to fit G\,+\,D$^{\prime}$, this results in an upshifted wide G band at $\sim$\,1600\,cm$^{-1}$.

In stage 2, the Raman spectrum evolves as follows~\cite{ferrari00}: a) the G peak position, denoted in literature as Pos(G) or $\omega_{\rm G}$, decreases from $\sim$\,1600\,cm$^{-1}$ towards $\sim$\,1510\,cm$^{-1}$; b) the TK relation fails and $I_{\rm D}/I_{\rm G}$ decreases towards 0; c) $\omega_{\rm G}$  becomes dispersive with the excitation laser energy, the dispersion increasing with disorder; d) there are no more well defined second-order peaks, but a small modulated bump from $\sim$\,2300\,cm$^{-1}$ to $\sim$\,3200\,cm$^{-1}$~\cite{martinsferreira10,ferrari00}.

In disordered carbons $\omega_{\rm G}$ increases as the excitation wavelength decreases, from IR to UV~\cite{ferrari00}. The dispersion rate, Disp(G)\,=\,$\Delta \omega_{\rm G}/\Delta E_{\rm L}$, increases with disorder. The G dispersion separates the materials into two types. In those with only $sp^{2}$ rings, Disp(G) saturates at $\sim$\,1600\,cm$^{-1}$, the G position at the end of stage 1. In contrast, for those containing $sp^{2}$ chains (such as in amorphous and diamond-like carbons), G continues to rise past 1600\,cm$^{-1}$ and can reach $\sim$\,1690\,cm$^{-1}$ for 229\,nm excitation~\cite{ferrari01,ferrari00}. On the other hand, D always disperses with excitation energy~\cite{ferrari00,ferrari01}. $\Gamma_{\rm G}$ always increases with disorder\cite{cn,cancado07,lucchese10,martinsferreira10}. Thus, combining $I_{\rm D}/I_{\rm G}$ and $\Gamma_{\rm G}$ allows to discriminate between stages 1 or 2, since samples in stage 1 and 2 could have the same $I_{\rm D}/I_{\rm G}$, but not the same $\Gamma_{\rm G}$, being this much bigger in stage 2\cite{cn,lucchese10,martinsferreira10}.

We note that Figure~\ref{f01} shows the loss of sharp second order features in the Raman spectrum obtained from the $L_{\rm D}$\,=\,2\,nm SLG. This is an evidence that the range of defect densities in our study covers stage 1 (samples with $L_{\rm D}$\,=\,24,\,14,\,13,\,7,\,5\,nm) and the onset of stage 2 (sample with $L_{\rm D}$\,=\,2\,nm).

\begin{figure*}[t!]
\centerline{\includegraphics[width=180mm]{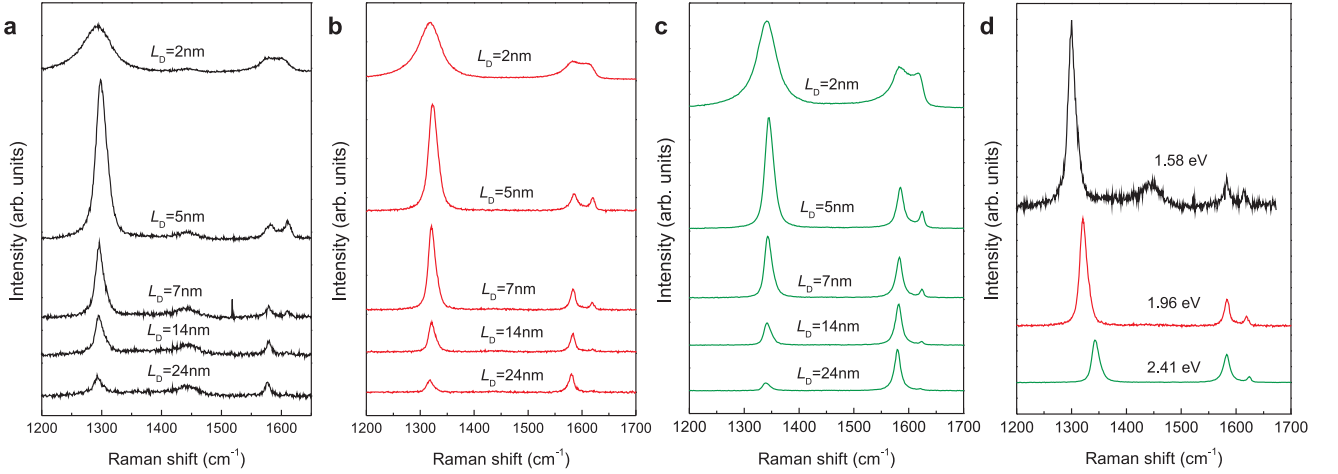}}
\caption{(a-c) Raman spectra of five distinct ion-bombarded graphene samples using the excitation laser energies (wavelengths) $E_{\rm L}\,=\,$1.58\,eV ($\lambda_{\rm L}\,=\,$785\,nm), $E_{\rm L}\,=\,$1.96\,eV ($\lambda_{\rm L}\,=\,$632.8\,nm), and $E_{\rm L}\,=\,$2.41\,eV ($\lambda_{\rm L}\,=\,$514.5\,nm), respectively. (d) Raman spectra of an ion-bombarded sample with $L_{\rm D}$\,=\,7\,nm obtained using these three excitation laser energies.\label{f02}}
\end{figure*}

Figures~\ref{f02}a-c report the first-order Raman spectra of our ion-bombarded SLGs measured at $E_{\rm L}\,=\,$1.58\,eV ($\lambda_{\rm L}\,=\,$785\,nm), 1.96\,eV (632.8\,nm),2.41\,eV (514.5\,nm), respectively. Figure~\ref{f02}d shows the Raman spectra of the ion-bombarded SLG with $L_{\rm D}$\,=\,7\,nm obtained using the three different laser energies. We note that $I_{\rm D}/I_{\rm G}$ considerably changes with the excitation energy. This is a well-know effect in the Raman scattering of $sp^{2}$ carbons\cite{mernagh,pocsik,ferrari01,ferrari00,cancado06,cancado07}. Ref.~\cite{cancado07} noted that the integrated areas of different peaks depend differently on excitation energy $E_{\rm L}$: while $A_{\rm D}$, $A_{\rm D^{\prime}}$, and $A_{\rm 2D}$ shown no $E_{\rm L}$-dependence, $A_{\rm G}$ was found to be proportional to $E_{\rm L}^4$. The independence of $A_{\rm 2D}$ on $E_{\rm L}$ agrees with the theoretical prediction~\cite{Basko08} if one assumes that the electronic scattering rate is proportional to the energy. However, a fully quantitative theory is not trivial since, in general, $A_{\rm D}$ depends not only on the concentration of defects, but on their type as well (e.g., only defects able to scatter electrons between the two valleys can contribute)~\cite{canedge,cedge,krauss2010}. Different defects can also produce different frequency and polarization dependence of $A_{\rm D}$~\cite{canedge,cedge,krauss2010}.

\begin{figure}[t!]
\centerline{\includegraphics[width=100mm]{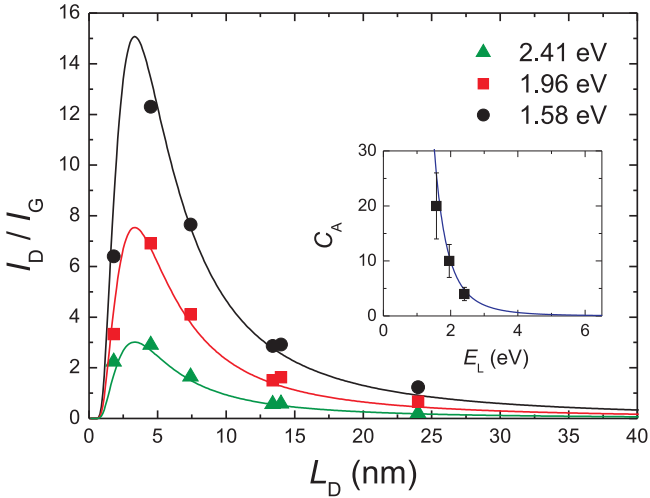}}
\caption{$I_{\rm D}/I_{\rm G}$ for all samples and laser energies considered here. Solid lines are fits according to equation~\ref{eq01} with $r_{\rm S}$\,=\,1\,nm, $C_{\rm S}$\,=\,0, and $r_{\rm A}$\,=\,3.1\,nm. The inset plots $C_{\rm A}$ as a function of $E_{\rm L}$. The solid curve is given by $C_{\rm A}\,=\,(160\pm48)\,E_{\rm L}^{-4}$.\label{f03}}
\end{figure}

Figure~\ref{f03} plots $I_{\rm D}/I_{\rm G}$ for all SLGs and laser energies. For all $E_{\rm L}$, $I_{\rm D}/I_{\rm G}$ increases as $L_{\rm D}$ decreases (stage 1), reaches a maximum at $L_{\rm D}$\,$\sim$\,3\,nm, and decreases towards zero for $L_{\rm D}$\,$<$\,3\,nm (stage 2). It is important to understand what the maximum of $I_{\rm D}/I_{\rm G}$ vs. $L_{\rm D}$ means. $I_{\rm D}$ will keep increasing until the contribution from each defect sums independently~\cite{lucchese10,cedge}. In this regime (stage 1) $I_{\rm D}$ is proportional to the total number of defects probed by the laser spot. For an average defect distance $L_{\rm D}$ and laser spot size $L_{\rm L}$, there are on average $(L_{\rm L}/L_{\rm D})^{2}$ defects in the area probed by the laser, thus $I_{\rm D}$\,$\propto\,(L_{\rm L}/L_{\rm D})^{2}$. On the other hand, $I_{\rm G}$ is proportional to the total area probed by the laser$\propto L_{\rm L}^{2}$, giving $I_{\rm D}/I_{\rm G}$\,$\propto\,1/L_{\rm D}^{2}$~\cite{lucchese10,ferrari00}. However, if two defects are closer than the average distance an e-h pair travels before scattering with a phonon, then their contributions will not sum independently anymore~\cite{lucchese10,martinsferreira10,beams,cedge}. This distance can be estimated as ${\emph v}_{\rm F}/\omega_{\rm D}\,\sim$\,3\,nm~\cite{cedge}, where ${\emph v}_{\rm F}\,\sim\,10^{6}$\,m/s is the Fermi velocity around the \textbf{K} and \textbf{K$^{\prime}$} points, in excellent agreement with the predictions of Refs.\cite{ferrari00} and the data of Refs.\cite{lucchese10,martinsferreira10,beams}. For an increasing number of defects (stage 2), where $L_{\rm D}$\,$<$\,3\,nm, $sp^2$ domains become smaller and the rings fewer and more distorted, until they open up. As the G peak is just related to the relative motion of $sp^2$ carbons, we can assume $I_{\rm G}$ roughly constant as a function of disorder. Thus, with the loss of $sp^2$ rings, $I_{\rm D}$ will decrease with respect to $I_{\rm G}$ and the $I_{\rm D}/I_{\rm G}$\,$\propto\,1/L_{\rm D}^{2}$ relation will no longer hold. In this regime, $I_{\rm D}/I_{\rm G}$\,$\propto\,M$ ($M$ being the number of ordered rings), and the development of a D peak indicates ordering, exactly the opposite to stage 1~\cite{ferrari00}. This leads to a new relation: $I_{\rm D}/I_{\rm G}$\,$\propto\,L_{\rm D}^{2}$~\cite{ferrari00}.

The solid lines in Fig.~\ref{f03} are fitting curves following the relation proposed in Ref.~\cite{lucchese10}:
\begin{equation}\label{eq01}
\frac{I_{\rm D}}{I_{\rm G}}\,=\,C_{\rm A}\frac{(r_{\rm A}^{2}-r_{\rm S}^{2})}{(r_{\rm A}^{2}-2r_{\rm S}^{2})} \left[e^{-\pi r_{\rm S}^{2}/L_{\rm D}^{2}}- e^{-\pi (r_{\rm A}^{2}-r_{\rm S}^{2})/L_{\rm D}^{2}}\right] .\;\;\;\;\;
\end{equation}
The parameters $r_{\rm A}$ and $r_{\rm S}$ are length scales which determine the region where the D band scattering takes place. $r_{\rm S}$ determines the radius of the structurally disordered area caused by the impact of an ion. $r_{\rm A}$ is defined as the radius of the area surrounding the point defect in which the D band scattering takes place, although the $sp^{2}$ hexagonal structure is preserved~\cite{lucchese10}. In short, the difference $r_{\rm A}\,-\,r_{\rm S}$ defines the Raman relaxation length of the D band scattering, and is associated with the coherence length of electrons which undergo inelastic scattering by optical phonons\cite{lucchese10,beams}. The fit in Figure~\ref{f02} is done considering $r_{\rm S}$\,=\,1\,nm (as determined in Ref.~\cite{lucchese10} and expected to be a structural parameter, i.\,e. $E_{\rm L}$ independent). Furthermore, within experimental accuracy, all data can be fit with the same $r_{\rm A}$\,=\,3.1\,nm, in excellent agreement with the values obtained in Refs.~\cite{lucchese10,martinsferreira10,beams}. Any uncertainty in $r_{\rm A}$ does not affect the results in the low defect density regime ($L_{\rm D}\,>\,10\,$nm) discussed later.

Ref.\cite{lucchese10} suggested that $I_{\rm D}/I_{\rm G}$ depends on both an activated (A) area, pounded by the parameter $C_{\rm A}$, and a structurally defective area (S), pounded by a parameter $C_{\rm S}$. Here we selected $C_{\rm S}\,=\,0$ in eq.~(\ref{eq01}) for two reasons: (i) $C_{\rm S}$ should be defect-structure dependent, and in the ideal case where the defect is the break-down of the C-C bonds, $C_{\rm S}$ should be null; (ii) here we do not focus on the large defect density regime, $L_{\rm D}\,<\,r{\rm _S}$. The parameter $C_{\rm A}$ in eq.~(\ref{eq01}) corresponds to the maximum possible $I_{\rm D}/I_{\rm G}$, which would be observed in the ideal situation where the D band would be activated in the entire sample with no break down of any hexagonal carbon ring~\cite{lucchese10}.

$C_{\rm A}$ has been addressed in Ref.~\cite{lucchese10} as related to the ratio between the scattering efficiency of optical graphene phonons evaluated between $\Gamma$ and \textbf{K}. As we show here, the large $I_{\rm D}/I_{\rm G}$ dependence on $E_{\rm L}$ comes from the change on $C_{\rm A}$, which suggests this parameter might also depend on interference effects, when summing the different electron/hole scattering processes that are possible when accounting for the Raman cross section\cite{janina2,NJP,Chen,Kalbac,venezuela2010}. Note that $C_{\rm A}$ decreases as the laser energy increases. The solid line in the inset to Fig.~\ref{f02} is the fit of the experimental data (dark squares) by using an empirical relation between the maximum value of $I_{\rm D}/I_{\rm G}$ and $E_{\rm L}$, of the form $C_{\rm A}\,=\,A\,E_{\rm L}^{-B}$. The fit yields $A\,=\,(160\pm48)$\,eV$^{4}$, by setting $B$\,=\,4 in agreement with Refs.\cite{cancado06,cancado07}.

We now focus on the low-defect density regime ($L_{\rm D}\,\geq\,10$\,nm), since this is the case of most interest in order to understand how Raman active defects limit the ultimate mobility of graphene samples~\cite{ni,kim,fuhrer}. In this regime, where $L_{\rm D}\,>\,2r_{\rm A}$, the total area contributing to the D band scattering is proportional to the number of point defects, giving rise to $I_{\rm D}/I_{\rm G}\,\propto\,1/L_{\rm D}^{2}$, as discussed above. For large values of $L_{\rm D}$, eq.~(\ref{eq01}) can be approximated to
\begin{equation}\label{eq11}
\frac{I_{\rm D}}{I_{\rm G}}\,\simeq\,C_{\rm A}\frac{\pi(r_{\rm A}^{2}-r_{\rm S}^{2})}{L_{\rm D}^{2}}\,.
\end{equation}
By taking $r_{\rm A}\,=\,3.1$\,nm, $r_{\rm S}$\,=\,1\,nm, and also the relation $C_{\rm A}\,=\,(160\pm48)E_{\rm L}^{-4}$ obtained from the fit of the experimental data shown in~Figure~\ref{f02}, eq.~(\ref{eq11}) can be rewritten as
\begin{equation}\label{eq12}
L_{\rm D}^{2}\,{\rm(nm^{2})}=\,\frac{(4.3\pm1.3)\times10^{3}}{E_{\rm L}^{4}}\left(\frac{I_{\rm D}}{I_{\rm G}}\right)^{-1}\,.
\end{equation} In terms of excitation laser wavelength $\lambda_{\rm L}$ (in nanometers), we have
\begin{equation}\label{eq13}
L_{\rm D}^{2}\,{\rm(nm^{2})}=\,\left(1.8\pm0.5\right)\times10^{-9}\lambda_{\rm L}^{4}\left(\frac{I_{\rm D}}{I_{\rm G}}\right)^{-1}\,.
\end{equation}
Equations (\ref{eq12}) and (\ref{eq13}) are valid for Raman data obtained from graphene samples with point defects separated by $L_{\rm D}\,\geq\,10$\,nm using excitation lines in the visible range. In terms of defect density $n_{\rm D}$(cm$^{-2}$)\,=\,10$^{14}$/$(\pi L_{\rm D}^{2})$, eqs. (\ref{eq12}) and (\ref{eq13}) become
\begin{equation}\label{eq14}
n_{\rm D}({\rm cm}^{-2})\,=\,(7.3\pm2.2)\times10^{9}E_{\rm L}^{4}\,\left(\frac{I_{\rm D}}{I_{\rm G}}\right)\,,
\end{equation}and
\begin{equation}\label{eq14}
n_{\rm D}({\rm cm}^{-2})\,=\,\frac{(1.8\pm0.5)\times10^{22}}{\lambda_{\rm L}^{4}}\,\left(\frac{I_{\rm D}}{I_{\rm G}}\right).
\end{equation}

\begin{figure}[t!]
\centerline{\includegraphics[width=100mm]{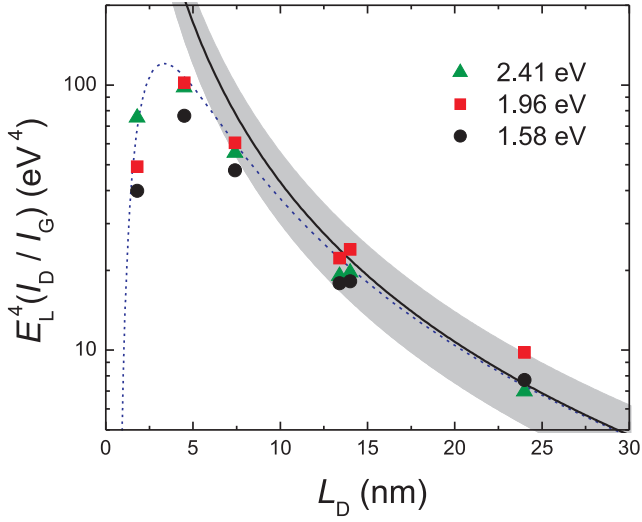}}
\caption{$E_{\rm L}^{4}(I_{\rm D}/I_{\rm G})$ as a function of $L_{\rm D}$ for the data shown in Figure~\ref{f02}. The dashed blue line is the plot obtained from the substitution of the relation $C_{\rm A}\,=\,(160)/E_{\rm L}^{-4}$ in equation~\ref{eq01}. The solid dark line is the plot of the product $E_{\rm L}^{4}(I_{\rm D}/I_{\rm G})$ as a function of $L_{\rm D}$ according to equation~\ref{eq12}. The shadow area accounts for the upper and lower limits given by the $\pm30\%$ experimental error.\label{f04}}
\end{figure}

Figure~\ref{f04} plots $E_{\rm L}^{4}(I_{\rm D}/I_{\rm G})$ as a function of $L_{\rm D}$ for the data shown in~Figure~\ref{f02}. The data with $L_{\rm D}\,>10$\,nm obtained with different laser energies collapse in the same curve. The dashed blue line is the plot obtained from the substitution of the relation $C_{\rm A}\,=\,(160)/E_{\rm L}^{4}$ in eq.~\ref{eq01}. The solid dark line is the plot $E_{\rm L}^{4}(I_{\rm D}/I_{\rm G})$ versus $L_{\rm D}$ according to eqs. (\ref{eq12}) and (\ref{eq13}). The shadow area accounts for the upper and lower limits given by the $\pm30\%$ experimental error. The plot in Fig.~\ref{f04} validates these relations for samples with $L_{\rm D}\,>\,10$\,nm.

\begin{figure}
\centerline{\includegraphics[width=90mm]{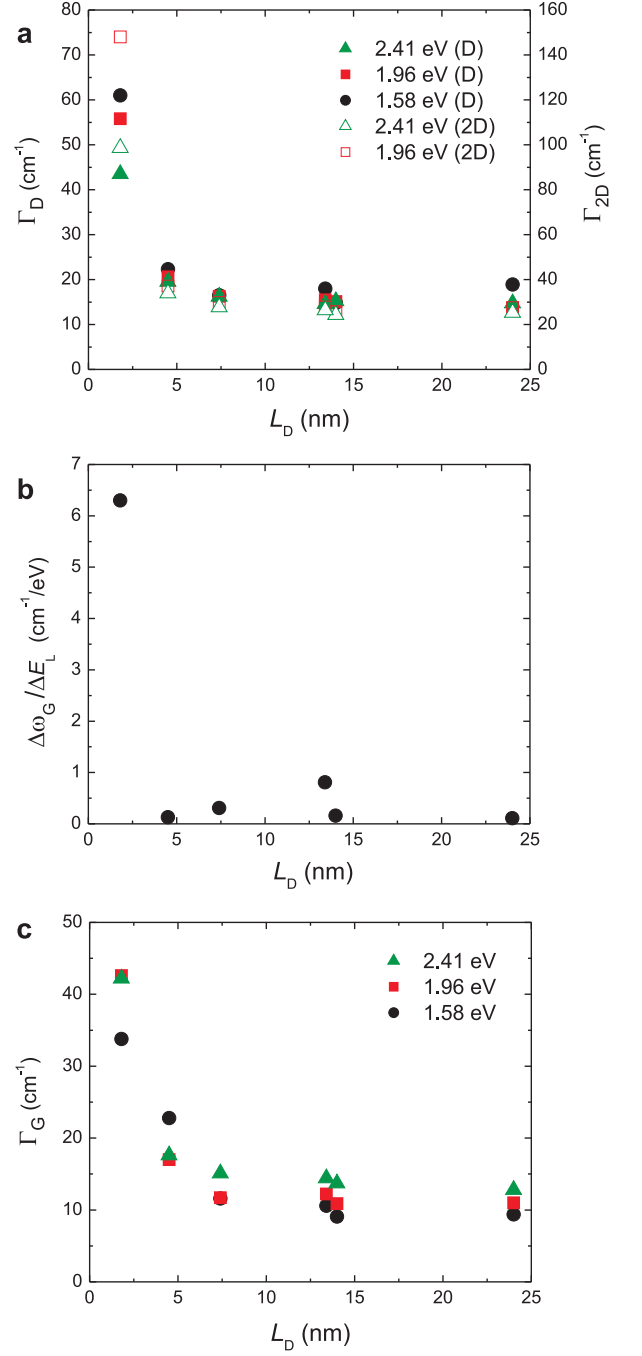}} \caption{(a) Plot of $\Gamma_{\rm D}$ and $\Gamma_{\rm 2D}$ versus $L_{\rm D}$. (b) G peak dispersion [Disp(G)\,=\,$\Delta \omega_{\rm G}/\Delta E_{\rm L}$] as a function of $L_{\rm D}$. $\Delta \omega_{\rm G}/\Delta E_{\rm L}$ remains zero until the onset of stage 2. (c) FWHM(G)\,=\,$\Gamma_{\rm G}$ as a function of $L_{\rm D}$. As suggested in Refs.\cite{cn,martinsferreira10}, $\Gamma_{\rm G}$ remains roughly constant until the onset of the second stage of amorphization, corresponding to the maximum $I_{\rm D}/I_{\rm G}$.\label{f05}}
\end{figure}

Figure~\ref{f05}a plots $\Gamma_{\rm D}$ and $\Gamma_{\rm 2D}$ as a function of $L_{\rm D}$. Within the experimental error, a dependence of $\Gamma_{\rm D}$ or $\Gamma_{\rm 2D}$ on the excitation energy during stage 1 can not be observed. D and 2D always disperse with excitation energy, with $\Delta \omega_{\rm D}/\Delta E_{\rm L}$\,$\sim$\,52\,cm$^{-1}/$eV, and $\Delta \omega_{\rm 2D}/\Delta E_{\rm L}$\,=\,2$\Delta \omega_{\rm D}/\Delta E_{\rm L}$.

Figures 5b,c plot the G peak dispersion Disp(G)$\,=\,\Delta\omega_{\rm G}/\Delta E_{\rm L}$ and $\Gamma_{\rm G}$\,=\,FWHM(G) as a function of $L_{\rm D}$, respectively. As shown in Figure~\ref{f05}b, $\Delta\omega_{\rm G}/\Delta E_{\rm L}$ remains zero until the onset of stage two, when it becomes slightly dispersive ($\Delta \omega_{\rm G}/\Delta E_{\rm L}$\,$\sim$\,6\,cm$^{-1}/$eV). $\Gamma_{\rm G}$ (Figure~\ref{f05}c) remains roughly constant at $\sim$\,14\,cm$^{-1}$, a typical value for as-prepared exfoliated graphene\cite{ferrari06,pisana,ferrarissc,lazzeri}, until the onset of stage 2 (corresponding to the maximum $I_{\rm D}/I_{\rm G}$) as suggested in Ref.~\cite{cn}, and shown in Ref.~\cite{martinsferreira10} for a single laser line $E_{\rm L}\,=\,$2.41\,eV. Combining $I_{\rm D}/I_{\rm G}$ and $\Gamma_{\rm G}$ allows to discriminate between stages 1 or 2, since samples in stage 1 and 2 could have the same $I_{\rm D}/I_{\rm G}$, but not the same $\Gamma_{\rm G}$, which is much larger in stage 2~\cite{cn,martinsferreira10}.

\section{Conclusions}

In summary, we discussed the use of Raman spectroscopy for quantifying the amount of point-like defects in graphene. We used different excitation laser lines in ion-bombarded samples in order to measure their respective $I_{\rm D}/I_{\rm G}$. We find that $I_{\rm D}/I_{\rm G}$, for a specific $L_{\rm D}$, depends on the laser energy. We presented a set of empirical relations that can be used to quantify point defects in graphene samples with $L_{\rm D}\,>\,$10\,nm via Raman spectroscopy using any laser line in the visible range. We show that the Raman coherence length $r_{\rm A}$ is $E_{\rm L}$-independent, while the strong $E_{\rm L}$ dependence for $I_{\rm D}/I_{\rm G}$ comes from the parameter $C_{\rm A}$.

\section{Acknowledgements}
We acknowledge funding from a Royal Society International Project Grant. ACF acknowledges funding from ERC grant NANOPOTS, EPSRC grant EP/G042357/1, a Royal Society Wolfson Research Merit Award, EU grants RODIN and Marie Curie ITN-GENIUS (PITN-GA-2010-264694), and Nokia Research Centre, Cambridge. LGC and AJ acknowledge the support from the Brazilian agencies CNPq and FAPEMIG. EHMF, FS, and CAA acknowledge financial support from Inmetro.

\end{document}